\documentclass[aps,prc,letterpaper,twoside,tightenlines,nofootinbib,showpacs
,preprint,superscriptaddress]{revtex4}
\usepackage{setspace}
\usepackage{epsfig,float}
\usepackage{amsmath}
\begin{document}

\newcommand{\Ima}{\textrm{Im}}
\newcommand{\Rea}{\textrm{Re}}
\newcommand{\mev}{\textrm{ MeV}}
\newcommand{\gev}{\textrm{ GeV}}
\newcommand{\rb}[1]{\raisebox{2.2ex}[0pt]{#1}}

\title{Two, three, many body systems involving mesons. Multimeson condensates}

\author{E. Oset}
\email{oset@ific.uv.es}
\affiliation{
Departamento de F\'{\i}sica Te\'orica and IFIC, Centro Mixto Universidad \\de Valencia-CSIC, Institutos de Investigaci\'on de Paterna, Apartado 22085, 46071 Valencia, Spain}

\author{M. Bayar}
\affiliation{Department of Physics, Kocaeli University, 41380 Izmit, Turkey}

\author{A. Dot\'e}
\affiliation{KEK Theory Center, Institute of Particle and Nuclear Studies (IPNS), High Energy Accelerator
Research Organization (KEK), 1-1 Oho, Tsukuba, Ibaraki, 305-0801, Japan}

\author{T. Hyodo}
\affiliation{Yukawa Institute for Theoretical Physics, Kyoto University, Kyoto 606-8502, Japan}

\author{K. P. Khemchandani}
\affiliation{Instituto de Fisica, Universidade de Sao Paulo, C.P. 66318, 05389-970 Sao Paulo, SP, Brazil}

\author{W. H. Liang}
\affiliation{Department of Physics, Guangxi Normal University, Guilin, 541004, People's Republic of China}

\author{A. Martinez Torres}
\affiliation{Instituto de Fisica, Universidade de Sao Paulo, C.P. 66318, 05389-970 Sao Paulo, SP, Brazil}

\author{M. Oka}
\affiliation{Department of Physics, Tokyo Institute of Technology, Meguro 152-8551, Japan; Advanced Science Research Center, Japan Atomic Energy Agency, Tokai, Ibaraki, 319-1195 Japan
}

\author{L. Roca}
\affiliation{Departamento de Fisica, Universidad de Murcia, E-30100 Murcia, Spain}

\author{T. Uchino}
\affiliation{
Departamento de F\'{\i}sica Te\'orica and IFIC, Centro Mixto Universidad \\de Valencia-CSIC, Institutos de Investigaci\'on de Paterna, Apartado 22085, 46071 Valencia, Spain}

\author{C.W. Xiao}
\affiliation{Institut f\"ur Kernphysik (Theorie), Institute for Advanced Simulation, and J\"ulich Center for Hadron Physics, Forschungszentrum J\"ulich,
D-52425 J\"ulich, Germany}

\date{\today}

\begin{abstract}

In this talk we review results from studies with unconventional many hadron systems containing mesons: systems with two mesons and one baryon, three mesons, some novel systems with two baryons and one meson, and finally systems with many vector mesons, up to six, with their spins aligned forming states of increasing spin. We show that in many cases one has experimental counterparts for the states found, while in some other cases they remain as predictions, which we suggest to be searched in BESIII, Belle, LHCb, FAIR and other facilities.

\end{abstract}


\maketitle

\section{Introduction}
   In this talk we review unconventional systems made by many hadrons, mostly mesons, or systems with some baryons and mesons, other than the also conventional mesonic atoms. The advent of the chiral unitary approach for meson meson interaction  \cite{npa,kaiser,markushin,juanito,rios} implementing unitarity in coupled channels from the basic interaction contained in the chiral Lagrangians \cite{Gasser:1983yg} has given rise to many states, found in poles of the scattering matrix. These states are known as dynamically generates states, kind of molecular states that arise from the interaction of the mesons and do not qualify as ordinary $q \bar q$ mesons, but are "extraordinary states" in the nomenclature used by Jaffe in the last Hadron Conference \cite{jaffe}. Similarly, the meson baryon interaction constructed implementing unitarity in coupled channels from the meson baryon chiral Lagrangians \cite{Ecker:1994gg,Bernard:1995dp} has given rise to many states that also qualify as dynamically generated states \cite{Kaiser:1995cy,
angels,ollerulf,Nieves:2001wt,Gamermann:2011mq,hyodo,ikeda,cola,Borasoy:2005ie,Oller:2005ig,
Borasoy:2006sr,Hyodo:2008xr}. An early review on these issues can be seen in \cite{review}. 
The generalization of the chiral Lagrangians to incorporate the interaction of vector mesons was also done in \cite{hidden1,hidden2,hidden4}. The unitarization of the vector-vector interaction in coupled channels using the information of \cite{hidden1,hidden2,hidden4} was also done in \cite{raquelvec}, with the surprise that some states emerged from the $\rho \rho$ interaction which could be associated to the $f_0(1370)$ and $f_2(1270)$. The generalization to SU(3) was done in \cite{gengvec} and 11 states were generated, which could be associated to known mesonic states. It was found there that the interaction in the spin $J=2$ channel was very strong, to the point that the  $f_2(1270)$ could be understood as a $\rho \rho$ molecular state.  

  The interaction for $\rho \rho$ in $J=2$ is so strong that one was lead to think that it would be possible to have states with many $\rho$ mesons with their spins aligned, such that all pairs would have $J=2$. With each of the pairs having $J=2$ in this case, the binding of the system was guaranteed. The question is then: how stable are these states? Unlike baryon many body systems where the conservation of baryonic number is responsible for the stability, for systems of many mesons one does not have meson number conservation and the multimeson states can decay into systems with fewer mesons. One might anticipate that these states would be highly unstable and the beautiful idea of the many meson systems would then be as short as the lifetime of these systems. However, it was found that this was not the case and in \cite{rocamulti} states up to six $\rho$ mesons were found with a width that made them observable. More surprising was the fact that the states found could be associated with known mesonic states. The exercise was repeated by studying meson systems with one $K^*$ and several $\rho$ mesons, and again relatively stable systems were found and associated to known $K^*$ states in \cite{rocayama}. By analogy, states with a $D^*$ and many $\rho$ mesons should also exist and predictions were done in \cite{dmulti}, but these states have not yet been experimentally investigated. 

  Apart from these states many other unconventional systems with three hadrons have been investigated and we shall report upon them in this review.

\section{Multirho states}

The standard tool to study three body systems are the Faddeev equations \cite{faddeev}, that, in spite of their formal simplicity, are rather involved technically and one sort or another of approximations is usually done to solve them numerically \cite{Alt:1967fx,epelbaum}. A different method, suited to the use of input from amplitudes obtained in the chiral unitary approach was done in \cite{MartinezTorres:2007sr,MartinezTorres:2008gy,Khemchandani:2008rk,MartinezTorres:2008kh}. Variational methods are also often used to study such systems \cite{Torres:2011jt,Jido:2008kp}.

  One of the approximations, which is often used is the Fixed Center Approximation (FCA) \cite{Chand:1962ec,Toker:1981zh,Barrett:1999cw,Deloff:1999gc,Kamalov:2000iy}. The method takes a cluster of two particles, which are bound and are supposed not to be much altered by the interaction with the third particle. Then, this third particle is allowed to interact multiply with the elements of the cluster. The amplitude for this multiple scattering is evaluated and then, eventually, bound states, or peaks, with a certain width if the system can decay, are obtained.  

\begin{figure}
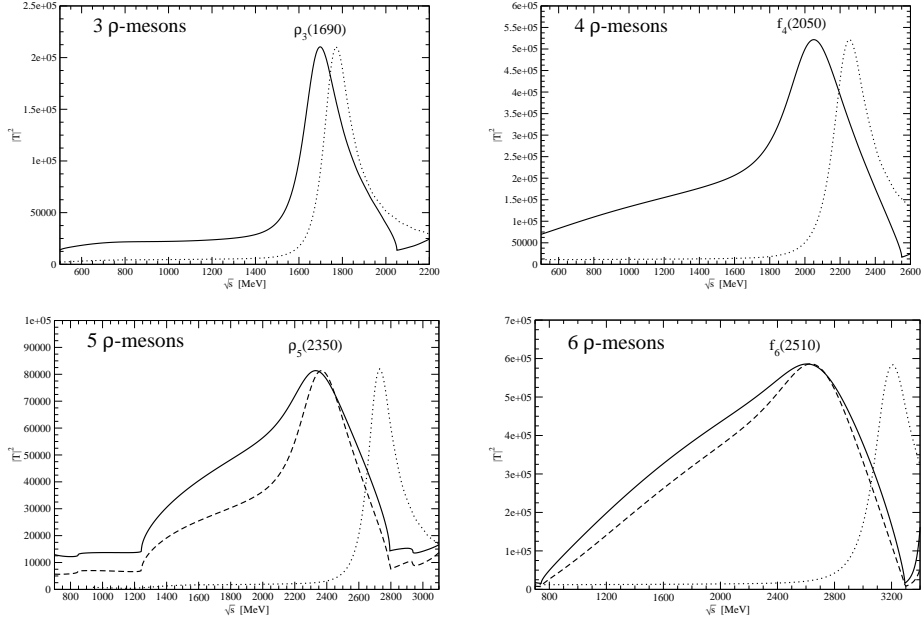

\begin{center}
\makebox[0pt]{\includegraphics[width=.35\linewidth]{figure5a.eps}\hspace{0.5cm}
\includegraphics[width=.35\linewidth]{figure5b.eps}}\\\vspace{0.3cm}
\makebox[0pt]{\includegraphics[width=.35\linewidth]{figure5c.eps}\hspace{0.5cm}
\includegraphics[width=.35\linewidth]{figure5d.eps}}
   \caption{Modulus squared of the unitarized multi-$\rho$ amplitudes.
   Dotted line: only single-scattering. Solid lines correspond to the prediction of the model. Dashed lines come from making a small change in a cut off. 
 The dashed and dotted lines have been normalized to
   the peak of the solid line for the sake of comparison 
   of the position 
   of the maxima}
     \label{fig:T2s}
\end{center}
\end{figure}

\begin{figure}[!t]
\begin{center}
\includegraphics[width=0.50\textwidth]{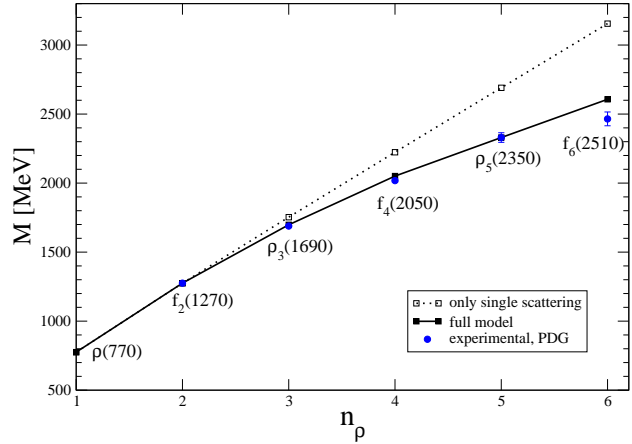}
\caption{Masses of the dynamically generated states as a function of the
number of constituent $\rho(770)$ mesons, $n_\rho$. Only single
scattering contribution (dotted line); full model (solid line);
experimental values from the PDG, (circles).
}
\label{fig:Mvsn}
\end{center}
\end{figure}

  Coming back to the multirho states, the work of \cite{rocamulti} proceeded as follows: two $\rho$ systems were allowed to interact in $J=2$, producing the $f_2(1270)$ state. Then a third $\rho$ meson was allowed to interact with this cluster, producing a $\rho$ state with $J=3$. Another $\rho$ meson was allowed to interact with this new cluster, which was made up a $\rho$ and a  $f_2(1270)$, producing a new state with isospin $I=0$ with $J=4$, and then the procedure was repeated iteratively till six $\rho$ mesons were put together and the width was still within measurable range. In this way six states were found that we plot in Figs. \ref{fig:T2s}, \ref{fig:Mvsn}. These states could be associated to the known states   $f_2(1270)$, $\rho_3(1690)$,   $f_4(2050)$,
 $\rho_5(2350)$ and   $f_6(2510)$. It should be stressed that there are no free parameters in the results of Figs. \ref{fig:T2s}, \ref{fig:Mvsn} for $\rho_3(1690)$,   $f_4(2050)$,
 $\rho_5(2350)$ and   $f_6(2510)$. The only free parameter in the theory was a cut off fitted to get the mass of the  $f_2(1270)$ in \cite{raquelvec}.

\section{$K^*$ multirho states} 

In a similar way to what is done with the multirho states, in \cite{rocayama} the interaction  of systems formed by a $K^*$ and many $\rho$ mesons was also studied and, once again, several states appeared which could be associated to the $K^*_2(1430)$, $K^*_3(1780)$, $K^*_4(2045)$, $K^*_5(2380)$. Another state, $K^*_6$, was also found, with a large width, but possibly identifiable as a meson state for which no experimental counterpart has been found yet.

 \section{$D^*$, multirho states}

The success of the two former studies suggested to also study states formed from one $D^*$ and many $\rho$ mesons.  The work was done in  \cite{dmulti}. The work was preceded by the study of the $D^* \rho$ interaction in \cite{nagavec}, where three $D$ states with spin J = 0, 1, 2
were obtained, the second one identified with the $D^*(2640)$
and the last one with the $D_2^*(2460)$. The first state, with J = 0,
was predicted at 2600 MeV with a width of about 100 MeV.
This state is also in agreement with the D(2600), which has a similar mass and
width, and which was reported experimentally after the theoretical work in \cite{delAmoSanchez:2010vq}.

   In \cite{dmulti} several states were also found with one $D^*$ and several $\rho$ mesons, all of them with their spins aligned to give states of increasingly larger spin. The states found in \cite{dmulti} were $D^*_3$, $D^*_4$, $D^*_5$ and $D^*_6$. However, unlike in the former cases, these states are not found in the list of the PDG \cite{pdg}. Their masses are predicted around $2800-2850\mev$, $3075-3200\mev$, $3360-3375\mev$ and $3775\mev$ respectively. And their widths are about $60-100\mev$, $200-400\mev$, $200-400\mev$ and $400\mev$ respectively. The existence of the analogous states discussed in the former sections  and the existence of  the $D$ states investigated in \cite{nagavec} give us much confidence that, with the time, this large spin $D$ states will also be found. 

\section{States with two mesons and one baryon}

These states were studied in \cite{MartinezTorres:2007sr,Khemchandani:2008rk}. We show them in Table \ref{table1} for states with strangeness S=-1. In the $S=0$ sector one finds several resonances, which are summarized in Table \ref{table2}. There is a  $N^*$ state around 1924 MeV, which is mostly $N K \bar{K}$. This state was first predicted in \cite{Jido:2008kp} using variational methods and corroborated in \cite{MartinezTorres:2008kh} using coupled channels Faddeev equations. In both works one  finds that the $ K \bar{K}$ pair is built mostly around the $f_0(980)$, but it also has a similar strength around the $a_0(980)$, both of which appear basically as a $ K \bar{K}$ molecule in the chiral unitary approach.

\begin{table}
\centering
\begin{tabular}{cccc}
\hline
&$\Gamma$ (PDG)&Peak position& $\Gamma$ (this work)\\
&(MeV)&(this work, MeV)& (MeV)\\
\hline
Isospin=1&&&\\
\hline
$\Sigma(1560)$&10-100&1590&70\\
$\Sigma(1620)$&10-100&1630&39\\
$\Sigma(1660)$&40-200&1656&30\\
$\Sigma(1770)$&60-100&1790&24\\
\hline
Isospin=0&&&\\
\hline
$\Lambda(1600)$&50-250&1568,1700&60-136\\
$\Lambda(1810)$&50-250&1740&20\\
\hline

\end{tabular}
\caption{$\Sigma$ and $\Lambda$ states obtained from the interaction of two mesons and one baryon.}\label{table1}
\end{table}

\begin{table}
\centering
\begin{tabular}{c|ccc|ccc}
\hline\hline
$I(J^P)$&\multicolumn{3}{c}{Theory}&\multicolumn{3}{c}{PDG data}\\
\hline
&channels&mass&width&name&mass&width\\
&&(MeV)&(MeV)&&(MeV)&(MeV)\\
\hline
$1/2(1/2^+)$&only $\pi\pi N$&1704&375&$N^*(1710)$&1680-1740&90-500\\
& $\pi\pi N$, $\pi K\Sigma$, $\pi K\Lambda$, $\pi\eta N$&$\sim$ no change&$\sim$ no change&&&\\
\hline
$1/2(1/2^+)$&only $\pi\pi N$&2100&250&$N^*(2100)$&1885-2270&80-400\\
& $\pi\pi N$, $\pi K\Sigma$, $\pi K\Lambda$, $\pi\eta N$&2080&54&&&\\
\hline
$3/2(1/2^+)$&$\pi\pi N$, $\pi K\Sigma$, $\pi K\Lambda$, $\pi\eta N$&2126&42&$\Delta(1910)$&1870-2152&190-270\\
\hline
$1/2(1/2^+)$&$N\pi\pi $, $N\pi\eta $, $NK\bar K$&1924&20&$N^*(?)$&?&?\\
\hline\hline
\end{tabular}
\caption{$N^{*}$ and $\Delta$ states obtained from the interaction of two mesons and one baryon.}\label{table2}
\end{table}

\section{Other three body states}
In \cite{Xiao:2011rc} the systems  $\bar{K}DN$, $NDK$ and $ND \bar{D}$ are investigated. Once more one finds quasibound states, relatively narrow, with energies $3150\mev, 3050\mev$ and $4400\mev$, respectively. All these states have $J^P=1/2^+$ and isospin $I=1/2$ and differ by their charm or strangeness content, $(S, C)=(-1,1), (1,1), (0,0),$ respectively.  The first state could perhaps be associated to the $\Xi_c(3123)$, which has unknown $J^P$, but the width obtained is a bit too large. The second state, of exotic nature, has no counterpart in the PDG. The third state is a regular $N^*$ state, but it contains hidden charm. One is making predictions that could be investigated in the coming Facilities of FAIR, or the BELLE upgrade, or the recently very successful LHCb. 

In \cite{Roca:2011br} pseudotensor mesons as three-body resonances are investigated. 
 One finds that the lightest pseudotensor mesons $J^{PC}=2^{-+}$ can be
regarded as molecules made of a pseudoscalar $(P)$ $0^{-+}$ and  a
tensor $2^{++}$ meson, where the latter is itself made of two vector
($V$) mesons. The author finds clear resonant structures which can be identified
with the  $\pi_2(1670)$, $\eta_2(1645)$ and $K^*_2(1770)$  ($2^{-+}$)
pseudotensor mesons. 

In \cite{Bayar:2013bta} the $\rho K \bar K$ system is studied and a quasibound state is found which is associated to the $\rho(1700)$. The $K \bar K$ system in this case clusters around the $f_0(980)$.

Similarly, in \cite{Bayar:2015oea}  the interaction of the a $\rho$ and $D^*$, $\bar D^*$ with spins aligned is studied using again the Fixed Center Approximation to the Faddeev equations. In this case an $I=1$ state with mass around 4340 MeV and narrow width of about 50 MeV is found.

 In \cite{Liang:2013yta} the $\eta K \bar K$ and $\eta' ′K \bar K$ systems are studied. The $\eta K \bar K$ is found to create some structure around 1490 MeV that could be identified with the $\eta(1475)$. However, such state was not found in a more detailed evaluation in  \cite{MartinezTorres:2011vh}. In this latter work, instead, some theoretical support was found for the $\pi(1300)$ and the recently claimed $f_0(1790)$ as molecular resonances made also of three hadrons. 
Coming back to the work of \cite{Liang:2013yta}, the  $\eta' K \bar K$ was also studied, but in this case only a cusp effect at threshold was found. 

In the case of two nucleons and one meson, the $DNN$ system was studied and quasi-bound states with isospin $I=1/2$ were found using two methods, the fixed center approximation to the Faddeev equation and the variational method approach to the effective one-channel Hamiltonian \cite{Bayar:2012dd}. It was found that the system had about  $\sqrt{s} \sim 3500$ MeV, bound by about 250 MeV from the $DNN$ threshold. Its width including both the mesonic decay and the $D$ absorption, was estimated to be about $20$-$40$ MeV. In this case, the $I=0$ $DN$ pair in the $DNN$ system was found to form a cluster similar to the $\Lambda_c(2595)$. It is remarkable that this system is more stable than its counterpart, the $\bar K NN$ system, where many theoretical studies coincide with having a larger width than the binding, that makes the experimental observation problematic (see a recent review on the subject in \cite{Kezerashvili:2015yza}).

\section{Conclusions}

   In this talk we have reported on studies of unconventional systems that have three hadrons, two mesons and a baryon, three mesons, two baryons and a meson, and many quasibound states were found which could be identified with known resonances. In other cases some predictions were done which could be tested in future experimental works. Particular interest was put in systems with many vector mesons, up to six mesons. We showed that the systems were very bound but they also decayed with larger widths as the number of mesons increases. We could show that in the case of multirho and $K^*$ multirho systems, the predicted states could be associated to already known resonances. In other cases, the states found, with high spin, remained as predictions that hopefully will be found in the future. 

\section{Acknowledgments}
  This work is partly supported by
the Spanish Ministerio de Economia y Competitividad and European FEDER funds under
the contract number FIS2011-28853-C02-01, FIS2011- 28853-C02-02, FIS2014-57026-REDT,
FIS2014-51948-C2- 1-P, and FIS2014-51948-C2-2-P, and the Generalitat Valenciana in the
program Prometeo II-2014/068. We
acknowledge the support of the European Community-Research Infrastructure Integrating
Activity Study of Strongly Interacting Matter (acronym HadronPhysics3, Grant Agreement
n. 283286) under the Seventh Framework Programme of EU.

\bibliographystyle{ursrt}

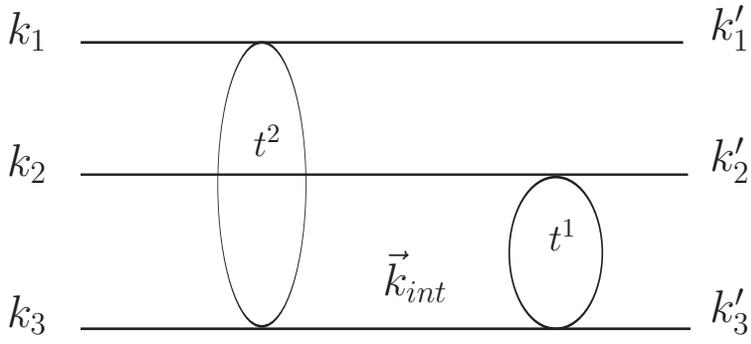
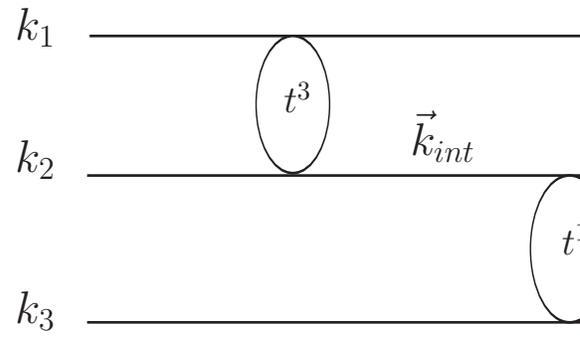
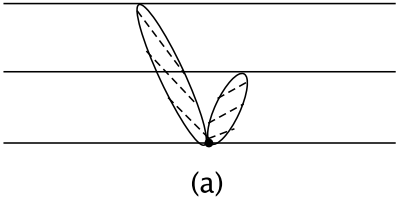
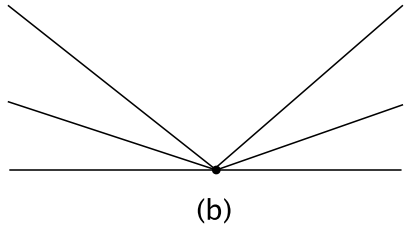
\end{document}